\documentclass[a4paper,11pt]{article}
\usepackage{mathrsfs} \usepackage{amsmath} \usepackage{amssymb} \usepackage{slashed}
\usepackage{graphicx} \usepackage{caption}
\begin{document}
\newcommand{\ba}{\begin{eqnarray}} \newcommand{\ea}{\end{eqnarray}}
\newcommand{\be}{\begin{equation}} \newcommand{\ee}{\end{equation}}
\renewcommand{\figurename}{Figure}
\renewcommand{\thefootnote}{\fnsymbol{footnote}}

\vspace*{1cm}
\begin{center}
 {\Large\textbf{A Model of Realistic Flavor Symmetry: Origin of Fermion Mass, Flavor Mixing and Leptogenesis}}

 \vspace{1cm}
 \textbf{Wei-Min Yang}

 \vspace{0.3cm}
 \emph{Department of Modern Physics, University of Science and Technology of China}

 \emph{Hefei 230026, People's Republic of China}

 \emph{E-mail: wmyang@ustc.edu.cn}
\end{center}

\vspace{1cm}
\noindent\textbf{Abstract}:
 I proposed a unified model of particle physic and cosmology in \cite {1}, which can simultaneously account for these origin of the inflation, dark energy, dark matter, neutrino mass and baryon asymmetry. I here focus on the fermion flavor issues in the unified model, which were not addressed previously. I introduce a realistic flavor symmetry to generate the fermion mass and mixing, from which we naturally derive the relationship between the quark mixing and the lepton mixing, and reveal the source of the difference between them. In particular, I derive a neutrino mass matrix which has a special structure form and only contains four parameters, but its numerical solutions are exactly accurately fitting to all the measured data of the neutrino mass spectrum and lepton mixing, and finely predict $m_{\nu_{2}}=0.01037$ eV. In addition, I discuss a new scenario of the leptogenesis in the model, which arises from two CP-asymmetric decays of a super-heavy neutral Dirac fermion, its CP asymmetry is closely related to the neutrino mass and mixing, via which we can correctly predict the baryon asymmetry. Lastly, I give several approaches to test the model. In short, the model can simply and elegantly account for the fermion flavor issues and the baryon asymmetry, and it has realistic and testable significance, therefore we expect the ongoing and future experiments to test the model.

\vspace{1cm}

\newpage
\noindent\textbf{I. Introduction}

\vspace{0.3cm}
 The standard model of particle physics (SM) have successfully accounted for numerous experimental measurements at the current day \cite{2}, nevertheless, it contains too many fermions: six flavors of the quarks and six flavors of the leptons, so there are a plethora of the flavor parameters related with them, furthermore, there are seemly no correlation among these flavor parameters. The origin of the fermion flavors is still a puzzle, and their implications for the universe evolution are also not understood, therefore, the fermion flavor implications have become the most mysterious aspect of the particle physics. At the present day, by means of plenty of analyses of the particle experimental data, the particle physicists have precisely measured the fermion mass and mixing parameters, which include the quark mixing matrix $U_{\text{CKM}}$ in the charged current couplings of the quarks \cite{3}, the lepton mixing matrix $U_{\text{PMNS}}$ in the charged current couplings of the leptons \cite{4}, and the two mass-squared differences in the neutrino oscillation \cite{5}, their current best-fit values are given as follows \cite{2},
\begin{alignat}{1}
 &U_{\text{CKM}}: \:\sin\theta^{q}_{12}\approx0.22501\,,\hspace{0.3cm} \sin\theta^{q}_{23}\approx0.04183\,,\hspace{0.3cm} 
    \sin\theta^{q}_{13}\approx0.003732\,,\hspace{0.3cm} \delta^{q}\approx0.3651\pi\,, \nonumber\\
 &U_{\text{PMNS}}: \:\sin\theta^{l}_{12}\approx0.554\,,\hspace{0.3cm} \sin\theta^{l}_{23}\approx0.747\,,\hspace{0.3cm} 
    \sin\theta^{l}_{13}\approx0.148\,,\hspace{0.3cm} \delta^{l}\approx1.19\,\pi\,, \nonumber\\
 & \triangle m^{2}_{21}\approx7.53\times10^{-5}\;\text{eV}^{2},\hspace{0.3cm}
    \triangle m^{2}_{32}\approx2.455\times10^{-3}\,\text{eV}^{2} (\text{NO})
    \;\text{or}\,-2.53\times10^{-3}\,\text{eV}^{2} (\text{IO}),
\end{alignat}
 where these notations are self-explanatory, $U_{\text{CKM}}$ and $U_{\text{PMNS}}$ are parameterized by the standard form, refer to the following Eq. (10). The NO and IO abbreviations respectively denote the normal and inverted order of the neutrino mass spectrum. Why do these two mixing matrices appear in the quark and lepton charged current couplings? What is the source of the great difference between them? Where is the source of the CP-violation phases in the mixing matrices? Is there really the underlying flavor symmetry for the fermions? or it is merely an imagined symmetry? Why the neutrino masses are far smaller than the charged lepton masses? The neutrino nature is a Dirac fermion or Majorana one? What mechanism is really responsible for generating the neutrino mass? In addition, what role does the fermion flavor play in the baryogenesis or leptogenesis? All of these issues make up the puzzles about the fermion flavor physics, they have therefore become the most challenging research in particle physics and cosmological phenomenology, and they are currently attracting more and more attentions of theoretical and experimental physicists \cite{6}.
 
 For decades, numerous proposals have been studied to address the above-mentioned issues of the fermion flavors. They include some extensions of the SM \cite{7}, some grand unified models \cite{8}, many special models of the fermion flavors \cite{9}, a great number of neutrino mixing models \cite{10}, a lot of mechanisms of neutrino mass generation \cite{11}, and some mechanisms of the baryogenesis and leptogenesis \cite{12}, and so on. Nevertheless, a wide variety of the proposals have a common shortcoming, namely they are only aiming at one or two specific aspects of the fermion flavor issues rather than considering internal connections among them, in particular, they lack the organic unification of particle physics and cosmology. In addition, some models of the fermion flavors are too unrealistic to believable, the suggested flavor symmetries are even an imaginary and unmeasurable thing, these theories are obviously inadvisable. In fact, the theoretical and experimental investigations of particle physics have clearly indicated that a realistic flavor model should be testable, each of the flavor parameters should be measurable and there are no the unphysical things. 

 Since the origin and evolution of the universe has the uniqueness, this destines that the particle phenomena are by no means isolated, but rather closely related to the cosmological phenomena, therefore, in the reference \cite{1} I proposed a new extension of the SM  based on the unification of particle physics and cosmology, which covers both the SM sector and the dark sector beyond the SM (BSM), this theory can successfully and simultaneously account for these phenomena of the inflation, dark energy, dark matter, neutrino mass and baryon asymmetry in a unified and integrated model. Here I will specially discuss the fermion flavor problems in the model, which were addressed previously. My starting point is based on such principle that the fermion flavor symmetry should be simple, realistic and measurable, the BSM flavor parameters are as few as possible, this is namely the Occam's Razor principle. In fact, the most realistic flavor symmetry is exactly the global flavor conservation of each flavor of the quarks and leptons, therefore I will explicitly define the fermion flavor states by their corresponding global flavor conservations so as to eliminate the vagueness of the fermion flavor, by which we can naturally and elegantly account for the origin of the fermion mass and mixing, in particular, generate neutrino mass and mixing, and further reveal the relationship and difference between the quark mixing and the lepton mixing. From the model Lagrangian I will derive the effective couplings of the neutrino and its mass matrix, furthermore, the neutrino mass matrix only contains four parameters, but its numerical solutions are exactly fitting to all the measured data in Eq. (1). In addition, I successfully implement a new leptogenesis scenario in the model, it is closely related to the neutrino mass and mixing and it can correctly predict the observed baryon asymmetry. Lastly, I will simply give a few of approaches for the model test. In short, this simple and reasonable model is very realistic and believable, therefore it is very worth studying in depth.

 The remainder of this paper is organized as follows. In Section II, I introduce the realistic flavor symmetry to generate the fermion mass and mixing, and then derive the neutrino mass matrix, and further find out its numerical solutions. In Section III, I discuss the new scenario of the leptogenesis. I simply give a few of proposals for the model test in Section IV. Section V is devoted to conclusions.

\vspace{0.6cm}
\noindent\textbf{II. Fermion mass and Mixing}

\vspace{0.3cm}
 The details of the particle model have been described in \cite{1}, here I only focus on the fermion flavor problem, specially discuss the fermion mass and mixing, which is originating from the following realistic flavor symmetry. It is well-known that the fermion flavor state is never explicitly defined in the SM, it is actually an ambiguous concept without a definitely physical identity. Before the electroweak symmetry breaking, the fermions are all massless, the fermion flavor states are both vague and undetectable because one can arbitrarily choose flavor basis (or rotate flavor space), this unphysical thing obviously conflicts with that a particle should have a definitely physical identity in the real universe, only after the electroweak symmetry breaking, the fermions become massive ones, the fermion flavor states are definitely identified as their mass eigenstates, which are namely the physical states identified and measured in all kinds of the particle experiments. Therefore, to overcome this shortcoming of the SM, we need explicitly define the fermion flavor state with definitely physical identity so as to eliminate the vagueness of the fermion flavor, this is perfectly solved by the following approach.
 
 First of all, the flavor state of each fermion species is explicitly defined by the unique physical identity which conserves its corresponding global flavor symmetry, there are namely
\begin{alignat}{1}
 &U(1)_{u}\otimes U(1)_{c}\otimes U(1)_{t}: \:q_{\alpha}=\left[\begin{matrix}u_{L\alpha}\\d'_{L\alpha}=U^{\text{CKM}}_{\alpha\beta}d_{L\beta}
   \end{matrix}\right],\hspace{0.3cm} u_{R\alpha}\,, \nonumber\\
 &U(1)_{d}\otimes U(1)_{s}\otimes U(1)_{b}: \:\tilde{q}_{\alpha}=\left[\begin{matrix}u'_{L\alpha}=U^{\dagger\text{CKM}}_{\alpha\beta}u_{L\beta}\\
   d_{L\alpha}\end{matrix}\right]=U^{\dagger\text{CKM}}_{\alpha\beta}q_{\beta}\,,\hspace{0.3cm} d_{R\alpha}\,, \nonumber\\
 &U(1)_{\nu_{1}}\otimes U(1)_{\nu_{2}}\otimes U(1)_{\nu_{3}}: \:l_{\alpha}=\left[\begin{matrix}
   \nu_{L\alpha}\\e'_{L\alpha}=U^{\text{CKM}}_{\alpha\beta}e_{L\beta}\end{matrix}\right],\hspace{0.3cm} \nu_{R\alpha}\,, \nonumber\\
 &U(1)_{e}\otimes U(1)_{\mu}\otimes U(1)_{\tau}: \:\tilde{l}_{\alpha}=\left[\begin{matrix}\nu'_{L\alpha}
   =U^{\dagger\text{CKM}}_{\alpha\beta}\nu_{L\beta}\\e_{L\alpha}\end{matrix}\right]=U^{\dagger\text{CKM}}_{\alpha\beta}l_{\beta},\hspace{0.3cm} e_{R\alpha}\,, \nonumber\\
 &U(1)_{\nu_{1}+\nu_{2}+\nu_{3}}: \:N_{L}\,,\hspace{0.3cm} N_{R}\,,
\end{alignat}
 where all kinds of the fermion notations are self-explanatory, refer to the description of Table 1 in \cite{1}. $[\alpha,\beta=1,2,3]$ are the generation (or family) indices, likewise, these subscripts of $[u,c,t]$, $[d,s,b]$, etc, are only employed to denote the meanings of the three generations, namely, they are corresponding to $\alpha=1,2,3$, one should not misunderstand them. The left-handed quark doublet have the Up-type flavor state $q$ and the Down-type flavor state $\tilde{q}$, they are related to each other by the unitary rotating $U_{\text{CKM}}$, their roles are very similar to that of the Higgs doublet $H$ and $\tilde{H}$ (see the following Eq. (3)). The left-handed $q$ and right-handed $u_{R}$ of the same generation have the same flavor number, so their three generations individually conserve the global $U(1)_{u}\otimes U(1)_{c}\otimes U(1)_{t}$. Likewise, the left-handed $\tilde{q}$ and right-handed $d_{R}$ of the same generation have the same flavor number, so their three generations individually conserve the global $U(1)_{d}\otimes U(1)_{s}\otimes U(1)_{b}$. Note that the flavor-rotating matrix $U_{\text{CKM}}$ is embedded in the definitions of the $q$ and $\tilde{q}$ structures from the very beginning, after the electroweak symmetry breaking, these components of $u_{L\alpha},u_{R\alpha}$ and $d_{L\alpha},d_{R\alpha}$ will automatically become the quark mass eigenstates (see the following Eq. (4)), and $U_{\text{CKM}}$ will also automatically appear in the quark charged current couplings (see the following Eq. (6)). Based on the unification of the quark flavor structure and the lepton one, and the minimum parameter principle, the flavor-rotating $U_{\text{CKM}}$ is in parallel embedded in the definitions of $l$ and $\tilde{l}$, so the parallel explanations apply to the lepton sector. Finally, both $\nu_{R\alpha}$ and $N_{L,R}$ are the neutral fermion singlets beyond the SM, but $\nu_{R\alpha}$ is the ultra-light right-handed neutrino and it is a dark neutrino with the ``$-1$" dark parity under the dark symmetry $Z_{2}^{\text{Dark}}$, in contrast, $N_{L,R}$ is a super-heavy neutral Dirac fermion without the dark parity, it is actually a mediator between the SM leptons and the dark $\nu_{R}$, so $N_{L,R}$ only conserves the total neutral lepton number of $U(1)_{\nu_{1}+\nu_{2}+\nu_{3}}$. In brief, there are the six quark flavor states and the six lepton flavor states, furthermore, each of the fermion flavor states includes the left-handed and right-handed components, they individually conserve their corresponding flavor number except the superheavy $N$.
 
 The scalar particles and their vacuum states in the model are given as follows,
\begin{alignat}{1}
 &H=\left[\begin{matrix}H^{+}\\H^{0}\end{matrix}\right],\hspace{0.3cm} \tilde{H}=i\tau_{2}H^{*},\hspace{0.3cm} 
   \Phi=\left[\begin{matrix}\Phi^{+}\\\Phi^{0}\end{matrix}\right],\hspace{0.3cm} \tilde{\Phi}=i\tau_{2}\Phi^{*},\hspace{0.3cm} 
   \phi=\phi^{*}, \nonumber\\
 &\langle H\rangle=\left[\begin{matrix}0\\\frac{v_{H}}{\sqrt{2}}\end{matrix}\right]\approx174\;\mathrm{GeV},\hspace{0.3cm} 
   \langle\Phi\rangle=\left[\begin{matrix}0\\\frac{v_{\Phi}}{\sqrt{2}}\end{matrix}\right]\sim1\;\mathrm{eV},\hspace{0.3cm} \langle\phi\rangle=v_{\phi}\sim10\;\mathrm{TeV}, \\
 &Z_{2}^{\text{Dark}}\xrightarrow{\langle\phi\rangle}\text{null}\,,\hspace{0.5cm}
   SU(2)_{L}\otimes U(1)_{Y}\xrightarrow{\langle H\rangle}U(1)_{\text{em}}\,, \nonumber
\end{alignat}
 where $\tau_{2}$ is the second Pauli matrix. The doublet $H$ is namely the SM Higgs field, while the doublet $\Phi$ and the real singlet $\phi$ are two dark scalar fields with the ``$-1$" dark parity under $Z_{2}^{\text{Dark}}$. $\langle\phi\rangle$ and $\langle H\rangle$ respectively break the dark $Z_{2}^{\text{Dark}}$ and the electroweak symmetry. $\Phi$ has inherently a superheavy mass of $M_{\Phi}\sim10^{11}$ GeV, so it only develops a tiny vacuum expectation value due to the see-saw mechanism. After the above vacuum breakings, $H^{0}$ and $\phi$ will respectively become the SM Higgs boson (whose mass has been measured as $M_{H}\approx125$ GeV) and the dark scalar boson (whose mass is estimated as $M_{\phi}\sim5$ TeV), while $\Phi$ still keeps its original structure due to $v_{\Phi}\ll M_{\Phi}$. In fact, $\Phi$ serves as the inflation field in \cite{1}, so it is absent at the low-energy scale. Note that all of the fermion flavor symmetries are unaffected by the above scalar vacuum breakings.

 Based on the above-mentioned particle contents and they following the gauge symmetry, dark parity and flavor symmetry, we can now write out the relevant Lagrangian in what follows. The Yukawa interactions are
\begin{alignat}{1}
 \mathscr{L}_{\text{Yukawa}}=&-\overline{\tilde{q}}\left[\begin{matrix}m_{d}&&\\&m_{s}&\\&&m_{b}\end{matrix}\right]d_{R}\frac{H\sqrt{2}}{v_{H}}
   -\overline{q}\left[\begin{matrix}m_{u}&&\\&m_{c}&\\&&m_{t}\end{matrix}\right]u_{R}\frac{\tilde{H}\sqrt{2}}{v_{H}} \nonumber\\
 &-\overline{\tilde{l}}\left[\begin{matrix}m_{e}&&\\&m_{\nu}&\\&&m_{\tau}\end{matrix}\right]e_{R}\frac{H\sqrt{2}}{v_{H}}
   -\overline{l}\left[\begin{matrix}y_{1}&&\\&y_{2}&\\&&y_{3}\end{matrix}\right]\nu_{R}\tilde{\Phi} \nonumber\\
 &-y_{4}\overline{l'}\left[\begin{matrix}1\\1\\1\end{matrix}\right]N_{R}\tilde{H}-y_{5}\overline{N_{L}}\,[1,1,1]\,\nu_{R}'\phi
   -M_{N}\overline{N_{L}}N_{R}+h.c.\,,
\end{alignat}
 where $l'$ and $\nu_{R}'$ are two new flavor states of the leptons, which are respectively related to the original $l$ and $\nu_{R}$ by the unitary rotating $T$ as follows,
\begin{alignat}{1}
 &l'=Tl\,,\hspace{0.5cm} \nu_{R}'=T\nu_{R}\,,\hspace{0.5cm} T=\frac{1}{\sqrt{65}} 
   \left[\begin{matrix}\frac{1+8\sqrt{2}}{\sqrt{3}}&\frac{-1+4\sqrt{2}}{\sqrt{6}}&\frac{1-4\sqrt{2}}{\sqrt{2}}\\
   \frac{1-4\sqrt{2}}{\sqrt{3}}&\frac{1+8\sqrt{2}-3\sqrt{65}}{2\sqrt{6}}&\frac{-1-8\sqrt{2}-\sqrt{65}}{2\sqrt{2}}\\
   \frac{1-4\sqrt{2}}{\sqrt{3}}&\frac{1+8\sqrt{2}+3\sqrt{65}}{2\sqrt{6}}&\frac{-1-8\sqrt{2}+\sqrt{65}}{2\sqrt{2}}\end{matrix}\right],\nonumber\\
 &T^{T}\left[\begin{matrix}1\\1\\1\end{matrix}\right]=\frac{12}{\sqrt{65}}\left[\begin{matrix}\frac{1}{4\sqrt{3}}\\\frac{1}{\sqrt{3}}\\-1\end{matrix}\right],\hspace{0.5cm} 
   [1,1,1]\,T=\frac{12}{\sqrt{65}}\,[\frac{1}{4\sqrt{3}}\,,\frac{1}{\sqrt{3}}\,,-1]\,.
\end{alignat}
 Obviously, the $N_{R}$ coupling to $l'$ and the $N_{L}$ coupling to $\nu_{R}'$ have all $S_{3}$ flavor symmetry of three generation permutation. The T matrix form in Eq. (5) is purely a phenomenological ansatz, but later it will be proved to be very successful. Those flavor conservations in Eq. (2) automatically constrain that these Yukawa couplings of the quarks and leptons are all diagonal matrices, so any non-diagonal element must be zero. After the electroweak breaking, the first three items of Eq. (4) directly gives rise to the quark and charged lepton masses, accordingly these left-handed components of $d_{L\alpha}$, $u_{L\alpha}$ and $e_{L\alpha}$ and their corresponding right-handed components automatically become the mass eigenstates. These quark and charged lepton masses in Eq. (4) as well as the $U_{\text{CKM}}$ elements in Eq. (6) are exactly the well-known flavor parameters in the SM, which have been precisely measured by the experiments. In contrast, the last four items of Eq. (4) belong to the flavor physics beyond the SM. The dark $\nu_{R}$ is only coupling to either $\tilde{\Phi}$ or $\phi$ because of the $Z_{2}^{\text{Dark}}$ constraint. In the last line of Eq. (4), the individual $U(1)_{\nu_{1}}\otimes U(1)_{\nu_{2}}\otimes U(1)_{\nu_{3}}$ are explicitly violated but the total $U(1)_{\nu_{1}+\nu_{2}+\nu_{3}}$ is still conserved, so any Majorana-type mass or couplings are all prohibited. These couplings of $y_{1,2,3}$, $y_{4,5}$ and $M_{N}$ are all the flavor parameters beyond the SM, they are responsible for the neutrino mass and mixing. We can reasonably assume that $y_{1,2,3}\sim10^{-2}$ is the same size as those Yukawa couplings of the charged lepton, while $y_{4,5}\sim10^{-4}$ is two weak couplings, so the neutrino flavor violation are relatively small terms. N has an inherent mass $M_{N}\sim10^{9}$ GeV, so the superheavy $N$ cannot appear at the low-energy scale. Finally, the quark and lepton masses and the coupling coefficients $y_{4,5}$ are all taken as real parameters since their complex phases can all be absorbed into the relevant fermion field redefinition, only the coupling coefficients $y_{1,2,3}$ remain complex parameters because their phases are irremovable further, thus the non-vanishing phases in $y_{1,2,3}$ can cause the CP violation in the Yukawa sector, simultaneously, they are namely the CP-violating source beyond the SM. Note that the model symmetries automatically guarantee the global conservations of both the baryon number and the lepton number.
 
 The gauge kinetic energy terms of the model Lagrangian contain the charged current couplings of the quarks as well as that of the leptons, by use of Eq. (2), we can directly give them as
\begin{alignat}{1}
 \mathscr{L}_{\text{Gauge}}\supset\, &-\frac{g}{\sqrt{2}}\,\overline{q}\,\gamma^{\mu}(W_{\mu}^{+}\tau^{+}+W_{\mu}^{-}\tau^{-})q
   =-\frac{g}{\sqrt{2}}\,\overline{\tilde{q}}\,\gamma^{\mu}(W_{\mu}^{+}\tau^{+}+W_{\mu}^{-}\tau^{-})\tilde{q} \nonumber\\
 &\xrightarrow{\text{EWB}}-\frac{g}{\sqrt{2}}\,[\overline{u_{L}}\gamma^{\mu}W_{\mu}^{+}U_{\text{CKM}}d_{L}
   +\overline{d_{L}}\gamma^{\mu}W_{\mu}^{-}U^{\dagger}_{\text{CKM}}u_{L}]\,, \nonumber\\
 &-\frac{g}{\sqrt{2}}\,\overline{l}\,\gamma^{\mu}(W_{\mu}^{+}\tau^{+}+W_{\mu}^{-}\tau^{-})l
   =-\frac{g}{\sqrt{2}}\,\overline{\tilde{l}}\,\gamma^{\mu}(W_{\mu}^{+}\tau^{+}+W_{\mu}^{-}\tau^{-})\tilde{l} \nonumber\\
 &\xrightarrow{\text{EWB}}-\frac{g}{\sqrt{2}}\,[\overline{\nu_{L}}\gamma^{\mu}W_{\mu}^{+}U_{\text{CKM}}e_{L}
   +\overline{e_{L}}\gamma^{\mu}W_{\mu}^{-}U^{\dagger}_{\text{CKM}}\nu_{L}]\,,
\end{alignat}
 where the notations are self-explanatory. Obviously, all the relations in Eq. (6) automatically follow those flavor conservations in Eq. (2), at the same time, the neutral current couplings (which are not written out here) of course keep those flavor conservations as well, therefore, the fermion flavor symmetries in Eq. (2) are actually obeyed by all of the gauge coupling terms. Note that $u_{L}$ and $d'_{L}=U_{\text{CKM}}d_{L}$ of the same generation have the same flavor number so that the three generation of them individually conserve $U(1)_{u}\otimes U(1)_{c}\otimes U(1)_{t}$, likewise, $d_{L}$ and $u'_{L}=U^{\dagger}_{\text{CKM}}u_{L}$ of the same generation have the same flavor number so that the three generation of them individually conserve $U(1)_{d}\otimes U(1)_{s}\otimes U(1)_{b}$. In brief, each of the gauge coupling terms in the quark sector can only conserve either $U(1)_{u}\otimes U(1)_{c}\otimes U(1)_{t}$ or $U(1)_{d}\otimes U(1)_{s}\otimes U(1)_{b}$, but it cannot simultaneously conserve them both. However, it should be stressed that before the electroweak breaking the $U_{\text{CKM}}$ matrix is hidden in the defined structures of the doublet $q$ and $\tilde{q}$, only after the electroweak breaking it can automatically expose in the charged current couplings of the quarks, but it never appears in the neutral current couplings, all of these results are purely derived from those arrangements of the fermion flavor states and favor symmetries in Eq. (2). The parallel discussions apply to the lepton gauge couplings, but the only difference between the lepton charged current couplings and the quark charged current couplings is that the neutrino flavor states $\nu_{L\alpha}$ in Eq. (6) are not aligned to its mass eigenstates because of the neutrino flavor-violating terms in Eq. (4), so now the $U^{\dagger}_{\text{CKM}}$ matrix in lepton sector cannot be identified as $U_{\text{PMNS}}$ as yet. Finally, we again stress that the complex phase in $U_{\text{CKM}}$ is the only CP-violating source in the quark sector, of course, it is also the only CP-violating source in the SM.
 
 The full scalar Lagrangian was given and discussed in \cite{1}, here I only write out the relevant part for the purpose of this paper, they are
\begin{alignat}{1}
\mathscr{L}_{\text{Scalar}}\supset\,\mu_{0}(H^{\dagger}\Phi\,\phi+\phi\,\Phi^{\dagger}H)-M^{2}_{\Phi}\,\Phi^{\dagger}\Phi\,,
\end{alignat}
 where $\mu_{0}\sim10^{7}$ GeV is the parameter of the triple scalar couplings with the mass dimension. The origin of $M_{N}$ and $\mu_{0}$ may theoretically arise from a special scalar symmetry breaking below the reheating temperature. In the reference \cite {1}, $\Phi$ serves as the inflation field, I in detail discussed its inflation evolution, and also obtained the inflationary potential form and the inflaton mass $M_{\Phi}\approx10^{11}$ GeV (which namely indicates the inflationary energy scale), at the end of the inflation the $\Phi$ decay can cause the reheating universe and the primordial plasma, in addition, I calculated out the reheating temperature $T_{\text{reh}}\approx5\times10^{10}$ GeV (which is a little below $M_{\Phi}$). In the current paper, however these terms in Eq. (7) have important implications for both the neutrino mass generation and the following leptogenesis.

 Because both $\Phi$ and $N$ are superheavy particles, which are absent at the low-energy scale, we can integrate out them in Eq. (4) by use of Eq. (7), then we can obtain the effective neutrino couplings, and further the neutrino mass matrix, they are given as follows,
\begin{alignat}{1}
 &\mathscr{L}^{\text{eff}}_{\text{neutrino}}=-\overline{l}\left(\frac{\mu_{0}}{M_{\Phi}^{2}}\left[\begin{matrix}y_{1}&&\\&y_{2}&\\&&y_{3}\end{matrix}\right]
   -\frac{y_{4}y_{5}}{M_{N}}\,T^{T}\left[\begin{matrix}1&1&1\\1&1&1\\1&1&1\end{matrix}\right]T\right)\nu_{R}\phi\tilde{H}+h.c. \nonumber\\
 &\hspace{1.5cm}=-\overline{l}\,M_{\nu}\,\nu_{R}\,\frac{\phi\tilde{H}\sqrt{2}}{v_{\phi}v_{H}}+h.c.\,,\hspace{0.5cm} M_{\nu}=M_{1}+M_{2}\,,\\
 &M_{1}=\frac{\mu_{0}v_{\phi}v_{H}}{\sqrt{2}M^{2}_{\Phi}}\left[\begin{matrix}y_{1}&&\\&y_{2}&\\&&y_{3}\end{matrix}\right]
   =\frac{v_{\Phi}}{\sqrt{2}}\left[\begin{matrix}y_{1}&&\\&y_{2}&\\&&y_{3}\end{matrix}\right]
   =v_{1}\left[\begin{matrix}ce^{i\varphi}&&\\&\frac{1}{\sqrt{2}}&\\&&1\end{matrix}\right], \nonumber\\
 &M_{2}=-\frac{y_{4}y_{5}v_{\phi}v_{H}}{\sqrt{2}M_{N}}\,T^{T}\left[\begin{matrix}1&1&1\\1&1&1\\1&1&1\end{matrix}\right]T
   =-v_{2}\left[\begin{matrix}\frac{1}{48}&\frac{1}{12}&-\frac{1}{4\sqrt{3}}\\\frac{1}{12}&\frac{1}{3}&-\frac{1}{\sqrt{3}}\\
   -\frac{1}{4\sqrt{3}}&-\frac{1}{\sqrt{3}}&1\end{matrix}\right], \nonumber
\end{alignat}
 where $v_{\Phi}=\frac{\mu_{0}v_{\phi}v_{H}}{M^{2}_{\Phi}}\sim1$ eV is namely the tiny vacuum expectation value of $\Phi$, and the relevant parameters are taken such as
\begin{alignat}{1}  
 &v_{H}\approx246\,\text{GeV},\hspace{0.2cm} v_{\phi}\sim10^{4}\,\text{GeV},\hspace{0.2cm} \mu_{0}\sim10^{7}\,\text{GeV},\hspace{0.2cm} 
   M_{N}\sim10^{9}\,\text{GeV},\hspace{0.2cm} M_{\Phi}\sim10^{11}\,\text{GeV}, \nonumber\\
 &[y_{1},y_{2},y_{3}]\sim10^{-2},\hspace{0.5cm} y_{4}\sim y_{5}\sim10^{-4}.
\end{alignat}
 In Eq. (8), the derived neutrino mass matrix has a special structure and form, namely $M_{\nu}$ consists of $M_{1}$ and $M_{2}$, furthermore, it is effectively characterized by only these four undetermined parameters of $[c,\varphi, v_{1},v_{2}]$, note that I have fixed $\frac{y_{2}}{y_{3}}=\frac{1}{\sqrt{2}}$ to minimize the parameters, which is justified by the later numerical results. Now $\varphi$ is the only remained complex phase and it is also the only CP-violating source beyond the SM. Later we will determine these parameters by fitting the measured data in Eq. (1). By use of Eq. (9), there are naturally $v_{1}\sim v_{2}\sim0.01$ eV, $c\lesssim1$ and $-\pi<\varphi<\pi$. When the effective neutrino couplings is compared with the quark and charged lepton Yukawa couplings in Eq. (4), the $M_{1}$ term follows the neutrino flavor conservation, which is similar to the quarks and charged lepton mass matrices, but it violates the CP symmetry due to the irremovable $\varphi$ phase, in contrast, the $M_{2}$ term explicitly violates the neutrino flavor conservation, but it hides the $S_{3}$ flavor symmetry, moreover, it conserves the total neutrino flavor number and the CP symmetry. As a result, this peculiar structure form of $M_{\nu}$ leads to the characteristic phenomena of the neutrinos and has an important implication for the leptogenesis. Obviously, this mechanism of generating neutrino mass and mixing is a new Dirac-type seesaw, which is very different from the traditional Majorana-type seesaw \cite{13}. 
 
 In Eq. (8), $M_{\nu}$ is explicitly a complex symmetric matrix, so it can be parameterized by three mass eigenvalues $m_{1,2,3}$ and one unitary matrix $U_{\nu}$, simultaneously, the neutrino flavor states are related to its mass eigenstates by the $U_{\nu}$ rotation, the detailed relations are given by
\begin{alignat}{1}
 &M_{\nu}=U_{\nu}\left[\begin{matrix}m_{1}e^{i\epsilon_{1}}&&\\&m_{2}e^{i\epsilon_{2}}&\\&&m_{3}e^{i\epsilon_{3}}\end{matrix}\right]U_{\nu}^{T},\hspace{0.3cm}
   \nu_{L}=U_{\nu}\,\nu_{L}^{\text{mass}},\hspace{0.3cm}
   \nu_{R}=U_{\nu}^{*}\left[\begin{matrix}e^{-i\epsilon_{1}}&&\\&e^{-i\epsilon_{2}}&\\&&e^{-i\epsilon_{3}}\end{matrix}\right]\nu_{R}^{\text{mass}},\nonumber\\
 &U_{\nu}=\left[\begin{matrix}e^{i\eta_{1}}&&\\&e^{i\eta_{2}}\approx1&\\&&1\end{matrix}\right]\left[\begin{matrix}
   c_{12}c_{13}&s_{12}c_{13}&s_{13}e^{-i\delta}\\-s_{_{12}}c_{23}-c_{12}s_{23}s_{13}e^{i\delta}&c_{_{12}}c_{23}-s_{12}s_{23}s_{13}e^{i\delta}&s_{23}c_{13}\\
   s_{_{12}}s_{23}-c_{12}c_{23}s_{13}e^{i\delta}&-c_{_{12}}s_{23}-s_{12}c_{23}s_{13}e^{i\delta}&c_{23}c_{13}\end{matrix}\right],
\end{alignat}
 where $U_{\nu}$ is parameterized by the standard form and these abbreviations are $s_{ij}=\sin\theta_{ij}$ and $c_{ij}=\cos\theta_{ij}$. In Eq. (10), these three phases of $\epsilon_{1,2,3}$ have no physical significance since they can be absorbed into the $\nu_{R}^{\text{mass}}$ redefinition, the physically meaningful quantities are actually these three mass eigenvalues and the $U_{\nu}$ matrix which contains three mixing angles and three phases, but $\eta_{2}$ is actually so small that it is negligible, see the following Table 1.
 
 Now we put $\nu_{L}=U_{\nu}\,\nu_{L}^{\text{mass}}$ into Eq. (6), namely the left-handed neutrino flavor states are rotated into its mass eigenstates, then the lepton mixing matrix in the charged current couplings of the leptons eventually becomes
\begin{alignat}{1}
 &U^{\dagger}_{\text{CKM}}U_{\nu}=\text{Diag}[e^{i\gamma_{1}},e^{i\gamma_{2}}\approx1,1]\,U_{\text{PMNS}}\,
   \text{Diag}[e^{i\gamma_{3}},e^{i\gamma_{4}},e^{i\gamma_{5}}\approx1]\,,\nonumber\\
 \text{or}\hspace{0.2cm}&U_{\nu}=U_{\text{CKM}}\,\text{Diag}[e^{i\gamma_{1}},e^{i\gamma_{2}}\approx1,1]\,U_{\text{PMNS}}\,
                \text{Diag}[e^{i\gamma_{3}},e^{i\gamma_{4}},e^{i\gamma_{5}}\approx1]\,,
\end{alignat}
 where $\gamma_{1},\ldots,\gamma_{5}$ are five phases associated with $U_{\text{PMNS}}$, nevertheless, $\gamma_{1,2}$ can be absorbed into the redefinitions of $e_{L,R}$ and $\gamma_{3,4,5}$ can be absorbed into the redefinitions of $\nu_{L,R}^{\text{mass}}$, so all of them are physically unmeasurable, but they mathematically have to exist to hold the above equality. Later the numerical calculations will concretely give $\gamma_{1}\approx0.1883\pi,\gamma_{2}\approx-6.1\times10^{-5}\pi,\gamma_{3}\approx-0.03282\pi,\gamma_{4}\approx0.01871\pi,\gamma_{5}\approx3.1\times10^{-5}\pi$. In Eq. (11), $U_{\nu}$ is equal to a product of these four matrix factors, this clearly shows a relationship between these three matrices of $U_{\nu}$, $U_{\text{CKM}}$ and $U_{\text{PMNS}}$, simultaneously, it also indicates the source of the great difference between the quark mixing and the lepton one. If there were no the superheavy $N$ in Eq. (4), then the neutrino flavor violation would be vanishing, the neutrino mass matrix would be as same as the quark and charged lepton cases, namely, there would be only the $M_{1}$ term in Eq. (8), while the $M_{2}$ term would be vanishing, thus there would be no the neutrino mixing (namely $U_{\nu}$ would be the unit matrix), then the lepton charged current couplings would become as same as the quark charged current couplings, there would be $U_{\text{PMNS}}=U^{\dagger}_{\text{CKM}}$, which is exactly shown in Eq. (6). In conclusion, the superheavy $N$ actually plays a key role in the neutrino mass and mixing, furthermore, it also plays another key role in the following leptogenesis.
 
 Next we need find out the values of these four parameters $[c,\varphi,v_{1},v_{2}]$ in $M_{\nu}$. Input a set of their values, according to Eq. (10), we can calculate out a solution of the neutrino mass spectrum $m_{1,2,3}$ and neutrino mixing matrix $U_{\nu}$, via Eq. (11) $U_{\nu}$ is further related to both the $U_{\text{CKM}}$ mixing and the $U_{\text{PMNS}}$ mixing. At present these ten quantities, namely, the two mass-squared differences of the neutrino, the four mixing parameters of $U_{\text{CKM}}$ and the four mixing parameters of $U_{\text{PMNS}}$ have been precisely measured, their values are all listed in Eq. (1). Therefore, we need find out the $M_{\nu}$ solutions which are fitting to all the data of Eq. (1), of course, this calculation is accomplished by a computer program. Note that these unphysical phases of $\epsilon_{1,2,3}$ and $\gamma_{1,2,3,4,5}$ are certainly involved in the computer program, but all of them are the output quantities rather than the input ones, since their values have no physical significance, so we only need fit the above-mentioned ten measured quantities.
 
 After make some efforts, we eventually find out the numerical solution of $M_{\nu}$ which are accurately fitting to all the measured data. The calculated results are in detail listed in Table 1.
\begin{table}
 \centering
 \begin{tabular}{|c|c|c|c|}
  \hline\hline
  \multicolumn{4}{|c|}{Input parameters of $M_{\nu}$} \\\hline
  $c$ &$\varphi$ &$v_{1}$(eV) &$v_{2}$(eV) \\\hline
  $0.043$ &$-0.387\pi$ &$0.0633$ &$0.0462$ \\\hline
 \end{tabular}
 \begin{tabular}{|c|c|c|c|c|c|c|c|c|}
  \hline
  \multicolumn{9}{|c|}{Output values of the neutrino mass spectrum and $U_{\nu}$ matrix} \\\hline
  $m_{2}$(eV) &$\triangle m^{2}_{21}$($\mathrm{eV}^{2}$) &$\triangle m^{2}_{32}$($\mathrm{eV}^{2}$) 
  &$\sin\theta^{\nu}_{12}$ &$\sin\theta^{\nu}_{23}$ &$\sin\theta^{\nu}_{13}$ &$\delta^{\nu}$ &$\eta_{1}$ &$\eta_{2}$ \\\hline
  $0.01037$ &$7.533\times10^{-5}$ &$2.455\times10^{-3}$ &$0.6554$ &$0.7803$ &$0.02312$ &$0.1908\pi$ &$0.1743\pi$ &$-1.3\times10^{-4}\pi$ \\\hline
  \end{tabular}
  \begin{tabular}{|c|c|c|c|c|c|c|c|}
  \hline
  \multicolumn{4}{|c|}{Fixed values of $U_{\text{CKM}}$}& \multicolumn{4}{|c|}{Output values of $U_{\text{PMNS}}$} \\\hline
  $\sin\theta^{q}_{12}$ &$\sin\theta^{q}_{23}$ &$\sin\theta^{q}_{13}$ &$\delta^{q}$
  &$\sin\theta^{l}_{12}$ &$\sin\theta^{l}_{23}$ &$\sin\theta^{l}_{13}$ &$\delta^{l}$ \\\hline
  $0.22501$ &$0.04183$ &$0.003732$ &$0.3651\pi$ &$0.5543$ &$0.7471$ &$0.1480$ &$1.191\pi$ \\\hline\hline
 \end{tabular}
 \caption{The numerical solutions of the neutrino mass spectrum and $U_{\nu}$ matrix, which are given by the $M_{\nu}$ model in Eq. (8). The $U_{\text{PMNS}}$ mixing is further predicted by Eq. (11). All of these results are accurately fitting to all the measured values in Eq. (1), and also finely predict the $m_{2}$ value.}
\end{table}
 The $c$ value is relatively small, so it has only a weak impact on the values of the neutrino mass and mixing, while the $\varphi$ value is in charge of the two phases of $\delta^{\nu}$ and $\eta_{1}$ in $U_{\nu}$, there would be $\eta_{1}=\delta^{\nu}=0$ and $\delta^{l}\approx\pi$ if $\varphi=0$. Both $v_{1}$ and $v_{2}$ are around $\sim0.05$ eV, this indicates that the $M_{1}$ contribution to $M_{\nu}$ and the $M_{2}$ one are the same size. This $M_{\nu}$ model predicts the normal order of the neutrino mass spectrum, moreover, it finely predicts $m_{2}=0.01037$ eV when $\triangle m^{2}_{21}$ and $\triangle m^{2}_{32}$ are exactly fitting to their measured values, note that the omitted $m_{1}$ and $m_{3}$ can be given by them. It can be seen from Table 1 that these three mixing angles of $U_{\nu}$ are predicted as $\theta^{\nu}_{12}\approx40.9^{\circ}, \theta^{\nu}_{23}\approx51.3^{\circ}, \theta^{\nu}_{13}\approx1.32^{\circ}$, which are close to the so-called bi-maximal pattern, in addition, the $\eta_{1}$ phase is not negligible but rather a key role in the successful fitting. Finally, this $U_{\nu}$ pattern guarantees that the left side of  Eq. (11) is equal to its right side, namely, it is accurately fitting to $U_{\text{CKM}}$ and $U_{\text{PMNS}}$.
 
 Finally, I stress that we only input these four effective parameters in $M_{\nu}$ without any fine-tuning, via the system of Eq. (10) and Eq. (11) we can accurately reproduce all the data in Eq. (1). When $U_{\text{CKM}}$ is fixed as its measured values, in fact, the model in total outputs seven measurable values, namely, the three masses $m_{1,2,3}$ and the four values of $U_{\text{PMNS}}$, among them, $\triangle m^{2}_{21}$, $\triangle m^{2}_{32}$ and $U_{\text{PMNS}}$ are all exactly fitting to their measured values, only $m_{2}=0.01037$ eV remains to be tested by future experimental measurements. In brief, the $M_{\nu}$ structure form in Eq. (8) can naturally lead that the neutrino mass and mixing are these results given by Table 1, via Eq. (11) the $U_{\nu}$ pattern can further guarantee that it are exactly fitting to the measured values of $U_{\text{CKM}}$ and $U_{\text{PMNS}}$. These discussions clearly show that the model has a stronger fitting power and predicting power, thereby this fully demonstrates that both the fermion flavor arrangements in Eq. (2) and the special rotation of the lepton flavors in Eq. (5) are scientific and reasonable, and this realistic solution for the fermion mass and mixing is successful and believable.
 
\vspace{0.6cm}
\noindent\textbf{III. Leptogenesis}

\vspace{0.3cm}
 In the paper of \cite{1}, I discussed that the inflation field decay $\Phi\rightarrow l^{c}+\nu_{R}$ can lead to the leptogenesis scenario, which only occurs in the reheating period after the inflation end, but this scenario is based on the previous premise of $\Gamma(\Phi\rightarrow l^{c}+\nu_{R})\ll\Gamma(\Phi\rightarrow H+\phi)$. In the current paper, a simple calculation gives $\Gamma(\Phi\rightarrow l^{c}+\nu_{R})/\Gamma(\Phi\rightarrow H+\phi)=(y_{\alpha}y_{\alpha}^{*})/(\frac{\mu_{0}}{M_{\Phi}})^{2}\gg1$ in the light of Eqs. (4), (7) and (9), so the original scenario is now invalid under the parameter selections of Eq. (9). Therefore, I here consider a new leptogenesis scenario, it directly arises from the two CP-asymmetric decays of $N\rightarrow l+H$ and $N\rightarrow\nu_{R}+\phi$, which can naturally occur in the primordial hot evolution after the universe reheating.
 
 According to Eq. (4), the $N$ decay have only the two channels of $N\rightarrow l+H$ and $N\rightarrow\nu_{R}+\phi$, moreover, their decay rates are of the same order of magnitude, Figure 1 draws their tree and one-loop diagrams.
\begin{figure}
 \centering
 \includegraphics[totalheight=6.5cm]{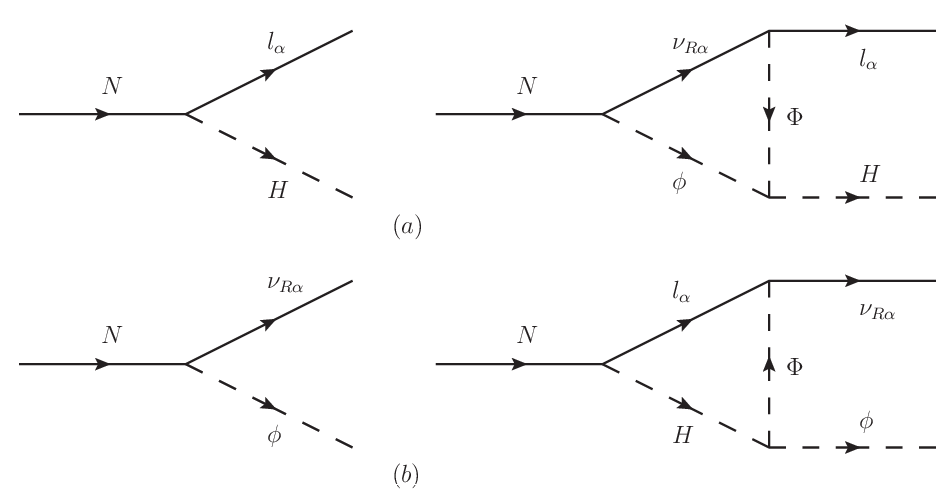}
 \caption{The tree and one-loop diagrams of $N\rightarrow l+H$ and $N\rightarrow\nu_{R}+\phi$. The CP asymmetry of $N\rightarrow l+H$ can generate the lepton asymmetry, while the CP asymmetry of $N\rightarrow\nu_{R}+\phi$ can generate the $\nu_{R}$ asymmetry, these two asymmetries are of the same amounts but opposite signs since the net total lepton number is conserved as zero. Below $T=M_{N}$, the $\nu_{R}$ is shortly decoupled from the primordial plasma, the $\nu_{R}$ asymmetry is forever frozen in the dark sector, whereas the lepton asymmetry in the SM sector is partly converted into the baryon asymmetry through the electroweak sphaleron transition.}
\end{figure}
 However, these two decays have the following three remarkable features. i) These two decays are out-of-equilibrium because their decay rates are smaller than the universe expansion rate. ii) Each decay rate has its own $CP$ asymmetry due to the interference between the tree amplitude and the one-loop one, the CP-violating source is exactly the $\varphi$ phase in $M_{\nu}$, as a result, these two decays can respectively lead to the lepton asymmetry and the $\nu_{R}$ asymmetry, nevertheless they are of the same amounts but opposite signs since the net total lepton number is conserved as zero in the universe. iii) At the same time, $\nu_{R}$ is shortly decoupled from the primordial plasma below $T=M_{N}$, thus the generated $\nu_{R}$ asymmetry are naturally separated from the lepton asymmetry, they respectively inhabit in the SM sector and the dark sector, so they cannot be erased each other. As a result, $\nu_{R}$ and its asymmetry are forever frozen in the dark sector, they are only diluted as the universe expansion, by contrast, the lepton asymmetry in the SM sector can partly be converted into the baryon asymmetry through the electroweak sphaleron transition \cite{14}. In short, this scenario is a new Dirac-type leptogenesis \cite{15a}, although it does not fully fulfil the Sakharov's three conditions \cite{15b}, it can be indeed put into effect on generating the matter-antimatter asymmetry in the universe.
 
 The relevant calculations of the $N$ decay are given as follows,
\begin{alignat}{1}
 &\Gamma_{N}=\Gamma(N\rightarrow l+H)+\Gamma(N\rightarrow\nu_{R}+\phi),\hspace{0.3cm}
   \Gamma_{\overline{N}}=\Gamma(N^{c}\rightarrow l^{c}+H^{c})+\Gamma(N^{c}\rightarrow\nu_{R}^{c}+\phi), \nonumber\\
 &\Gamma_{N}=\Gamma_{\overline{N}}\approx\frac{3M_{N}}{32\pi}(2y_{4}^{2}+y_{5}^{2})
   =(2\frac{y_{4}}{y_{5}}+\frac{y_{5}}{y_{4}})\frac{\sqrt{2}\,\text{Tr}[-M_{2}]M_{N}^{2}}{32\pi\,v_{\phi}v_{H}}
   <H=\frac{1.66\sqrt{g_{*}}\,T^{2}}{M_{\text{Pl}}}|_{T=M_{N}}\,, \nonumber\\
 &A_{\text{CP}}=\frac{\Gamma(N\rightarrow l+H)-\Gamma(N^{c}\rightarrow l^{c}+H^{c})}{\Gamma_{N}}
   =-\frac{\Gamma(N\rightarrow\nu_{R}+\phi)-\Gamma(N^{c}\rightarrow\nu_{R}^{c}+\phi)}{\Gamma_{N}} \nonumber\\
 &\hspace{0.75cm}=\frac{M_{\Phi}^{2}\text{Im}\text{Tr}[M_{2}M_{1}^{*}]}{3\pi(2y_{4}^{2}+y_{5}^{2})(v_{\phi}v_{H})^{2}}
   f(\frac{M^{2}_{N}}{M^{2}_{\Phi}})
   =\frac{\frac{\mu_{0}}{M_{N}}\,\text{Im}\text{Tr}[M_{2}M_{1}^{*}]}
     {\sqrt{2}\,\pi(2\frac{y_{4}}{y_{5}}+\frac{y_{5}}{y_{4}})v_{\Phi}\text{Tr}[-M_{2}]}
   f(\frac{M^{2}_{N}}{M^{2}_{\Phi}})\,,\\
 &f(\frac{M^{2}_{N}}{M^{2}_{\Phi}})=f(x)=1+\frac{\ln(1+x)}{x}=2-\frac{x}{2}+\frac{x^{2}}{3}-\frac{x^{3}}{4}+\cdots\approx2\,,\nonumber
\end{alignat}
 where $ f(\frac{M^{2}_{N}}{M^{2}_{\Phi}})$ is derived from the loop integration, in addition, I make use of $M_{1}$ and $M_{2}$ in Eq. (8) to express the above relations, which are substitute for the Yukawa coupling coefficients. $M_{\text{Pl}}=1.22\times10^{19}$ GeV is the Planck mass, the effective number of relativistic degrees of freedom is $g_{*}=113$ at $T=M_{N}$ in the model. From Eq. (12) we can clearly see that this leptogenesis scenario is closely related to the neutrino mass and mixing. From the foregoing discussions there are $\frac{y_{4}}{y_{5}}\sim1$, $\text{Tr}[-M_{2}]\approx0.063$ eV and $v_{\phi}v_{H}\sim2.5\times10^{6}$ $\text{GeV}^{2}$, then there is naturally $\Gamma_{N}<H$, note that this result is independent of the $M_{N}$ value, so the $N$ decay is indeed out-of-equilibrium. From the numerical solutions in Table 1 we can calculate out $\text{Im}\text{Tr}[M_{2}M_{1}^{*}]/\text{Tr}[-M_{2}]\approx-3.9\times10^{-5}$ eV, since there are $ v_{\Phi}\sim1$ eV, $\frac{\mu_{0}}{M_{N}}\sim10^{-2}$ and $\frac{y_{4}}{y_{5}}\sim1$, then we can naturally obtain $A_{\text{CP}}\sim-6\times10^{-8}$, note that this value and its negative sign together are very vital for producing correctly the baryon asymmetry. The source of the CP-asymmetry is from the $\varphi$ phase in $M_{1}$, which is namely the only CP-violating phase beyond the SM, it has nothing to do with the CP-violating phase of $U_{\text{CKM}}$ in the SM. Finally, we point out that the dilute process of $l^{c}+\nu_{R}\rightarrow\phi+H$ via the t-channel $N$ mediation is always invalid at the any temperature because its reaction rate is severely out-of-equilibrium by a simple calculation. 
 
 After the $N$ decays are completed, as a result, these two CP asymmetries in Eq. (12) can respectively generate asymmetric lepton and asymmetric $\nu_{R}$, they are the same amounts but opposite signs, so the net total lepton number is still conserved as zero. On the other hand, below $T=M_{N}$ the dark $\nu_{R}$ is essentially decoupling from the primordial plasma, thus $\nu_{R}$ and its asymmetry together are forever frozen in the dark sector, they are only diluted as the universe expanding. From the entropy conservation, we can derive that in later period the $\nu_{R}$ effective temperature is $T^{3}_{\nu_{R}}/T^{3}\approx0.0363$ (whereas the $\nu_{L}$ effective temperature is $T^{3}_{\nu_{L}}/T^{3}=\frac{4}{11}$). In the SM sector, the lepton asymmetry is partly converted into the baryon asymmetry through the electroweak sphaleron transition which conserves the $B-L$ number. Therefore, the present-day baryon asymmetry is simply calculated as follows \cite{16},
\begin{alignat}{1}
 &\eta_{B}=Y_{B}(T_{0})[\frac{s}{n_{\gamma}}]_{T_{0}}=Y_{B}(T_{\text{ew}})[\frac{s}{n_{\gamma}}]_{T_{0}}
   =c_{s}Y_{B-L}(T_{\text{ew}})[\frac{s}{n_{\gamma}}]_{T_{0}}=c_{s}Y_{B-L}(M_{N})[\frac{s}{n_{\gamma}}]_{T_{0}}\,, \nonumber\\
 &\text{at} \:T=M_{N},\hspace{0.2cm} Y_{B-L}=-\frac{n_{l}-\overline{n}_{l}}{s}=\frac{n_{\nu_{R}}-\overline{n}_{\nu_{R}}}{s}
  =-\frac{n_{N}A_{\text{CP}}}{s}\approx-0.42\,\frac{A_{\text{CP}}}{g_{*}}\,, 
\end{alignat}
 where $T_{0}$ denotes the present-day universe temperature, $Y_{B}(T)=[\frac{n_{B}-\overline{n}_{B}}{s}]_{T}$ is the yield, $s$ is the entropy density, $[\frac{s}{n_{\gamma}}]_{T_{0}}\approx7.38$ (because the $\nu_{R}$ contribution leads to $g_{*}(T_{0})\approx4.1$), $c_{s}=\frac{28}{79}$ is the SM sphaleron coefficient, and $g_{*}=113$ at $T=M_{N}$. If $A_{\text{CP}}\approx-6.32\times10^{-8}$ is given by Eq. (12), then put it into Eq. (13), thus we can accurately predict $\eta_{B}\approx6.14\times10^{-10}$, which is exactly fitting to the best-fit value observed by the current multiple experiments \cite{17}. By the way, the number density $n_{N}$ is suppressed exponentially as the universe temperature is falling, so the main bulk of $Y_{B-L}$ is essentially generated around $T=M_{N}$. If one uses Boltzmann equations to solve numerically $Y_{B-L}$, then the obtained result is actually the same order of magnitude as the one given by Eq. (13), only the prefactor coefficients of them are slightly different, but this difference is not significant because we can always compensate it by another choice of the $A_{\text{CP}}$ value.
 
 Finally, we can further derive that at the present day there is $[n_{\nu_{R}}+\overline{n}_{\nu_{R}}]/n_{\gamma}=\frac{3}{4}\times0.0363\approx0.0272$, while the ratio of $[n_{\nu_{R}}-\overline{n}_{\nu_{R}}]/[n_{\nu_{R}}+\overline{n}_{\nu_{R}}]\approx-A_{\text{CP}}$ is so far maintained, therefore there is $[n_{\nu_{R}}-\overline{n}_{\nu_{R}}]/n_{\gamma}\approx1.72\times10^{-9}$,  which is about three times $\eta_{B}$, so the $\nu_{R}$ asymmetry can indeed be ignored in the current universe. In addition, we can calculate out the effective number of the neutrinos at the recombination as $N_{\text{eff}}=3[1+(T_{\nu_{R}}/T_{\nu_{L}})^{4}]\approx3.14$, which is also safely within the current limit given by the CMB data analysis \cite{18}. In a word, all of the results of the model are self-consistent and very well in agreement with the present measurements.

\vspace{0.6cm}
\noindent\textbf{IV. Model Test}

\vspace{0.3cm}
 Finally, I simply give a few of proposals for the model test. Once the neutrino mass spectrum are eventually determined by the ongoing and future neutrino experiments, they are then compared with the predictions in Table 1, this will directly test the model. In the model, the phenomena of the BSM flavor physics arise from those couplings of the new particles, for instance, the leptogenesis discussed in the last Section. The effects of the BSM flavor physics can also occur in the low-energy processes, for example, the lepton flavor violation $\mu^{-}\rightarrow e^{-}+\gamma$, a new contribution to the anomalous magnetic moment of the muon, as shown the (a) diagram in Fig. 2, the pair annihilation of Higgs bosons into a pair of the dark right-handed neutrinos, namely $H+\bar{H}\rightarrow\nu_{R}+\nu_{R}^{c}$\,, as shown the (b) diagram in Fig. 2, etc, these processes are all detectable and measurable at the current laboratories and LHC \cite{19}.
\begin{figure}
 \centering
 \includegraphics[totalheight=6cm]{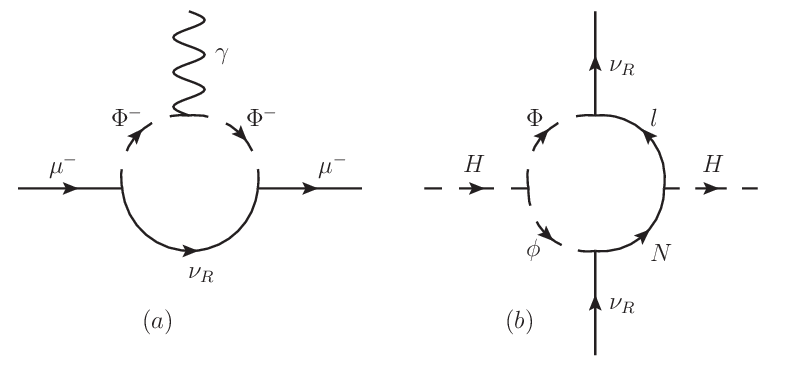}
 \caption{The (a) diagram makes a BSM contribution to the anomalous magnetic moment of the muon. The (b) diagram indicates the pair annihilation of Higgs bosons into a pair of the dark right-handed neutrinos, namely $H+\bar{H}\rightarrow\nu_{R}+\nu_{R}^{c}$\,.}
\end{figure}
 From the model couplings and the determined flavor parameters, we can calculate these processes and give their theoretical predictions, by comparison between the measured data and these predictions, we can reliably test the model. Of course, the more detailed discussions about the model test are beyond the focus of the current paper, it is more suitable that they are left for further study.

\vspace{0.6cm}
\noindent\textbf{V. Conclusions}

\vspace{0.3cm}
  Based on the new extension of the SM suggested in \cite{1}, in the current paper I specially discuss the flavor issues of the fermion mass and mixing in the model, which were not addressed previously. First of all, I explicitly define the fermion flavor states and their corresponding flavor symmetries in Eq. (2), this thus eliminates the vagueness of the fermion flavor state as well as the unphysical flavor parameters, and guarantees that each of the fermion states as well as each of the flavor parameters are physical significance and measurable, there are no extra and unphysical things. Next I in detail discuss all kinds of the fermion flavor conservations in the Yukawa couplings as well as the gauge couplings. However, the neutrino is an exceptional case, its flavor conservation is explicitly violated by their couplings to the superheavy $N$, but these flavor-violating couplings hide the $S_{3}$ permutation symmetry, from which the neutrino mixing is exactly arising, and eventually the neutrino mixing makes the difference between the quark mixing and the lepton mixing through Eq. (11). In Eq. (8) I derive the effective neutrino couplings at the low energy and obtain the neutrino mass matrix with the special structure form. $M_{\nu}$ is only determined by the four effective parameters, but its numerical solutions are accurately fitting to all the measured data in Eq. (1) and finely predict $m_{2}=0.01037$ eV, evidently, the $M_{\nu}$ model shows a stronger predicting power. 
  
  In addition, I discuss the new scenario of the leptogenesis in the model, which arises from the two CP-asymmetric decays in Figure 1, these two decays can respectively lead to the lepton asymmetry and the $\nu_{R}$ asymmetry with the same amounts but opposite signs, but these two asymmetries are shortly separated from each other as $\nu_{R}$ is decoupling from the hot bath below $T=M_{N}$, the former is partly converted into the baryon asymmetry through the SM sphaleron transition, whereas the latter is forever frozen in the dark sector. In particular, this leptogenesis scenario is closely related to the neutrino mass and mixing by Eq. (12), from which we can correctly predict the observed baryon asymmetry. Lastly, I suggest some approaches to test the model, which show that the BSM flavor physics has important implications for the particle phenomena and early universe evolution. In short, all of these discussions clearly demonstrate that the simple and realistic flavor model is very successful and believable.

  In summary, this paper supplements a detailed discussion about the fermion flavor issues for the particle model of \cite{1}, thus they together make up a more complete theory. This new theory covers both the SM physics (visible sector) and the BSM one (dark sector), moreover, it can simultaneously and excellently account for the fermion flavor issues, the matter-antimatter asymmetry, and some cosmological phenomena (which includes the inflation, dark matter and dark energy, see \cite{1}), of course, all aspects of the model are self-consistent and very well in agreement with the experimental data. Therefore, we believe that the unified model has surely scientific and realistic significance, it is very worth studying in depth, and we expect the ongoing and future experiments to test it.

\vspace{0.6cm}
 \noindent\textbf{Acknowledgements}

\vspace{0.3cm}
 I would like to thank my wife for her great helps. This research is supported by the Fundamental Research Funds for the Central Universities of China under Grant No. WY2030040065.

\end{document}